\documentclass[12pt]{iopart}
\usepackage{amssymb}
\usepackage{graphicx,bm}
\usepackage{setstack, cite}
\usepackage{lscape}
\begin{document}

\title[]{A report on the nonlinear squeezed states and their non-classical properties
of a generalized isotonic oscillator}

\author{V~Chithiika Ruby and  M~Senthilvelan$^{\dagger}$}

\address{Centre for Nonlinear Dynamics, School of Physics,
Bharathidasan University, Tiruchirappalli - 620 024, India.}
\ead{velan@cnld.bdu.ac.in}

\begin{abstract}
We construct nonlinear squeezed states of a generalized isotonic oscillator potential.
We demonstrate the non-existence  of dual counterpart of
nonlinear squeezed states in this system. We investigate statistical properties exhibited by
the squeezed states, in particular Mandel's parameter,
second-order correlation function, photon number distributions and parameter $A_3$ in detail. We also examine the
quadrature and amplitude-squared squeezing effects. Finally, we derive expression for the
$s$-parameterized quasi-probability distribution function of these states. All these information about the system are
new to the literature.
\end{abstract}

\maketitle

\section{Introduction}
Very recently studies have been made to analyze the generalized isotonic oscillator potential,
$V(y) = \left(\frac{m_0 \omega^2 }{2} y^2 + \frac{g_a(y^2 - a^2)}{(y^2 + a^2)^2}\right)$, in
different perspectives \cite{car,fellows, berger,sesma, hall1, hall2,sen, chi1,chi2,proy,quesne_iso}. The associated
Schr\"{o}dinger equation can be written as (after suitable rescaling)
\begin{eqnarray}
-\frac{1}{2}\frac{d^2 \psi_n(x)}{dx^2}+ \frac{1}{2}\left(x^2 + \frac{8(2x^2 - 1)}{(2x^2 +1)^2}\right)
\psi_n(x) =  E_n \psi_n(x).
\label{se}
\end{eqnarray}

Equation (\ref{se}) admits eigenfunctions and energy eigenvalues as  \cite{car}
\begin{eqnarray}
\psi_n(x) &=& {\cal N}_n \displaystyle{\frac{{\cal P}_n(x)}{(1+2x^2)}} e^{\displaystyle-{x^2}/2},
\label{eigfn}\\
E_n &=& -\frac{3}{2} + n, \quad \qquad \qquad n = 0,3,4,5,...,
\label{energy}
\end{eqnarray}
where the polynomial factors ${\cal P}_n(x)$ are given by
\begin{eqnarray}
\fl \qquad \qquad {\cal P}_n(x) = \left\{\begin{array}{c}
\hspace{-1.2cm} 1, \quad \qquad \qquad \qquad \qquad \qquad \qquad \qquad\quad \mbox{if}\;\;n = 0 \\
  H_n(x) + 4 n H_{n-2}(x) + 4 n (n-3) H_{n-4}(x), \;\mbox{if}\;\;n = 3, 4, 5,...\\
  \end{array}\right.
\end{eqnarray}
and the normalization constant
\begin{eqnarray}
{\cal N}_n =\left[\displaystyle{\frac{(n-1)(n-2)}{2^n n! \sqrt{\pi}}}\right]^{1/2}, \quad n = 0,3,4,5,....
\end{eqnarray}
We consider (\ref{se}) as the number operator equation after subtracting
the ground state energy ${\displaystyle E_0 = -\frac{3}{2}}$ from it, that is
\begin{eqnarray}
\hat{N}_0|n\rangle = n|n\rangle.
\label{no}
\end{eqnarray}

In a very recent paper \cite{chi2}, we have addressed the method of finding the
deformed ladder operators $\hat{N}_{-}$ and $\hat{N}_{+}$ from the solution (\ref{eigfn}).
The deformed ladder operators $\hat{N}_{-}$ and $\hat{N}_{+}$ satisfy the relations \cite{chi2}
\begin{eqnarray}
\hat{N}_{-}|n\rangle &=& \sqrt{n}\;f(n)\;|n-1\rangle, \label{lada} \\
\hat{N}_{+}|n\rangle &=& \sqrt{n+1}\;f(n+1)\; |n+1\rangle,
\label{lad10}
\end{eqnarray}
with $f(n) = \sqrt{(n-1)(n-3)}$. Since $f(n)$ has zeros at $n = 1$ and $3$, we relate  
the annihilation ($\hat{a}$) and creation operators ($\hat{a}^{\dagger}$) 
to the deformed ladder operators $\hat{N}_{-}$ and $\hat{N}_{+}$ through the relations,
\begin{eqnarray}
\hat{a} = \frac{1}{f(\hat{N}_0 + 1)} \hat{N}_{-}, \qquad
\hat{a}^{\dagger} =\frac{1}{f(\hat{N}_0)} \hat{N}_{+} , \quad n = 0, 3, 4, 5,... 
\label{lad10a}
\end{eqnarray} 
in which we preserve the ordering of operators $f(\hat{N}_{0}),\; \hat{N}_{-}$ and $\hat{N}_{+}$. 
Specifically the operators $\hat{a}$ and $\hat{a}^{\dagger}$ act 
on the states $|0\rangle$ and $|3\rangle$ yield 
\begin{eqnarray}
\hat{a}|0\rangle &=& \frac{1}{f(\hat{N}_{0}+1)} \hat{N}_{-}|0\rangle = 0, \qquad  
\hat{a}^{\dagger}|0\rangle = \frac{1}{f(\hat{N}_{0})} \hat{N}_{+}|0\rangle = 0, \\
\hat{a}|3\rangle &=&  \frac{1}{f(\hat{N}_{0}+1)} \hat{N}_{-}|3\rangle =  0, \qquad  
\hat{a}^{\dagger}|3\rangle =  \frac{1}{f(\hat{N}_{0})} \hat{N}_{+}|3\rangle = \sqrt{4}\;|4\rangle.
\label{lad10c}
\end{eqnarray} 
For the remaining states, the operators produce 
\begin{eqnarray}
\hat{a}|n\rangle &=& \sqrt{n}\;|n-1\rangle, \\
\hat{a}^{\dagger}|n\rangle &=& \sqrt{n+1}\;|n+1\rangle, \quad n = 4, 5, 6, 7, ...
\label{lad10b}
\end{eqnarray}
and $\hat{N}_{0} = \hat{a}^{\dagger} \hat{a}$. 

Since $\hat{N}_{-} |0\rangle = 0$ and $\hat{N}_{+}|0\rangle = 0$,
the ground state can be considered as an isolated one.  Further, the expression $\hat{N}_{-}|3\rangle = 0$ implies
that the first excited state $|3\rangle$ acts as a
ground state. This is due to the reason that $f(n)$ has
zeros at $n = 1$ and $3$. Because of this fact, the Hilbert space ${\cal H}$
consists of states $|0\rangle, |3\rangle, |4\rangle,...$ splits up into two invariant
sub-spaces, namely (i) $|\Psi\rangle = |0\rangle$ and (ii) $|\Psi^{'}\rangle = \sum^{\infty}_{n = 3} c_{n} |n\rangle$ for the
operators $\hat{N}_{-}$ and $\hat{N}_{+}$  \cite{Manko}. We consider the sub-Hilbert space, ${\cal H'}$, spanned
by the eigenstates, $|3\rangle, |4\rangle, |5\rangle,...$ and exclude the ground state $|0\rangle$ for further discussion.

The operators $\{\hat{N}_{-}, \hat{N}_{+},\hat{N}_{0}\}$ satisfy the following deformed
$su(1,1)$ algebra
\cite{ proy, Bambah, quadratic}
\begin{eqnarray}
[\hat{N}_{+},\hat{N}_{-}]|n\rangle = [5\hat{N}_{0} - 3 \hat{N}^2_{0}]|n\rangle, \qquad \;
[\hat{N}_{0}, \hat{N}_{\pm}]|n\rangle = \pm \hat{N}_{\pm} |n\rangle
\label{lad11}
\end{eqnarray}
with Casimir operator of the type \cite{quesne1993}
\begin{eqnarray}
\hat{C} = \hat{N}_{-}\hat{N}_{+} + h(\hat{N}_0) = \hat{N}_{+}\hat{N}_{-} + h(\hat{N}_0 - 1),
\label{lad12}
\end{eqnarray}
where $h(\hat{N}_0)$ is a real function which is of the form \cite{quesne1993}
\begin{eqnarray}
h(\hat{N}_0) = \frac{5}{2}\hat{N}_0(\hat{N}_0+ 1)- \hat{N}_0 (\hat{N}_0+1)(\hat{N}_0 + \frac{1}{2}).
\end{eqnarray}

We note here that a physical interpretation for the deformed operators 
was already given in Refs. \cite{Manko, Dodo}. In the present case also, 
we observe that the frequency of vibrations of the nonlinear oscillator 
depends on the energy of vibrations. To demonstrate this let us consider the Hamiltonian  
$\tilde{H} = \frac{1}{2}\left(\hat{N}_{+}\hat{N}_{-} + \hat{N}_{-}\hat{N}_{+}\right)$ 
associated with the quantum $f$-deformed nonlinear oscillator. The 
energy eigenvalues in the  Fock space is then given by $\tilde{E}_{n} = \frac{1}{2}[n (1-5n+2n^2)]$ 
\cite{Manko, Dodo}. The Heisenberg equation of motion for $\hat{N}_{-}$ (or $\hat{N_{+}}$) now reads 
\begin{eqnarray}
\dot{\hat{N}}_{-} + i [\hat{N}_{-}, \tilde{H}(\hat{N}_{0})] = 0 \;\Rightarrow\;\dot{\hat{N}}_{-} + i \omega_{\pm}(\hat{N}_0)\hat{N}_{-} = 0, 
\label{hes}
\end{eqnarray}
where $\omega_{+}(\hat{N}_0) = \tilde{H}(\hat{N}_{0}+1)-\tilde{H}(\hat{N}_0) = 3 \hat{N}^2_{0} - 2 \hat{N}_{0} - 1$, 
$\omega_{-}(\hat{N}_0) = \tilde{H}(\hat{N}_{0})-\tilde{H}(\hat{N}_0-1) = 3 \hat{N}^2_{0} - 4 \hat{N}_{0} + 2$ and 
the square bracket denotes the usual commutator. In terms of the evolution operator, 
$U(t) = e^{i \tilde{H}(\hat{N}_{0})(t - t_0)}$, the solution to (\ref{hes}) can be written as 
\begin{equation}
\hat{N}_{-}(t) = e^{-i \omega_{+}(\hat{N}_{0})(t -t_0)}\hat{N}_{-}(t_0). 
\label{eqo}
\end{equation}
Expression (\ref{eqo}) shows that the quantum $f$-oscillator vibrates with a frequency 
depends on the energy $\tilde{E}_{n}$.

The aim of this paper is to construct the nonlinear squeezed states of the
system (\ref{se}). A squeezed state is one of the minimum uncertainty states in which the
fluctuation of one photon-quadrature component is less than the quantum limit \cite{walls}. This can be achieved
by increasing or decreasing one of the photon-quadrature dispersions in such a way that the Heisenberg
uncertainty relation is not violated \cite{squeez_op, yuen, squeezing, dar}. Squeezed states can be
produced by acting with the squeezing operator
$S(\xi) = \exp{\left(\frac{\xi}{2} \hat{a}^{\dagger^2} - \frac{\xi^*}{2} \hat{a}^{2} \right)}$
on the coherent state or ground state or first order excited state of a quantum system, where $\hat{a}$ and $\hat{a}^{\dagger}$ are
annihilation and creation operators respectively and $\xi$ is
a complex parameter. The method of constructing nonlinear squeezed 
states in the $su(1,1)$ algebra was discussed in Ref. \cite{oba}.  The nonlinear squeezed states \cite{kwek} have applications in
quantum cryptography \cite{crypto}, quantum teleportation \cite{tele} and moreover they have also been proposed
for high precision measurements such as improving the sensitivity of Ramsey fringe interferometry \cite{ram} .
During the past three decades considerable efforts have been made towards the methods of generating
squeezed states in particular optical four-wave mixing and optical fibers, parametric amplifiers,
non-degenerate parametric oscillators and so on \cite{yuen, mix, fibre, amplifier, pa_oscillator}.

Motivated by these recent developments we intend to construct nonlinear squeezed
states for the generalized isotonic
oscillator potential. By transforming the deformed ladder operators suitably we identify
the Heisenberg algebra and the squeezing operators. While one of the operators
produces nonlinear squeezed states the other one fails to produce another set of
nonlinear squeezed states (dual pair) \cite{dual}.
Besides constructing nonlinear squeezed states we also
investigate the non-classical properties exhibited by the
nonlinear squeezed states, by investigating Mandel's parameter, second-order
correlation function and parameter $A_3$.  We examine non-classical nature
of the states by evaluating quadrature squeezing and amplitude-squared squeezing.
Further, we derive analytical expressions for the $s$-parameterized function
for the non-classical states.  The
partial negativity of the $s$-parameterized function confirm the non-classical properties
of the nonlinear squeezed states. All these informations about the
system (\ref{se}) are new to the literature.

We organize our presentation as follows. In the following section, we discuss the method of obtaining Heisenberg algebra
from the deformed annihilation and creation operators.
In section 3, we construct nonlinear squeezed states from the Heisenberg algebra
for this nonlinear oscillator. Consequently, we analyze certain photon statistical properties,
normal quadrature squeezing and amplitude-squared squeezing properties exhibited by the
nonlinear squeezed states and the harmonic oscillator squeezed
states in section 4. Followed by this, in section 5, we study quadrature distribution and quasi-probability
distribution function for the dual pairs of nonlinear squeezed states.
Finally, we present our conclusions in section 6.

\section{\bf Deformed oscillator algebra and transformations \cite{chi3}}
\label{sec2}
To construct nonlinear squeezed states \cite{kwek, Gbook} of (\ref{se}), we
transform $\hat{N}_{-}$ or/and $\hat{N}_{+}$ suitably, such a way that the 
newly transformed operators satisfy the Heisenberg algebra.
We consider all three possibilities in the following.

First let us rescale $\hat{N}_{+}$ as \cite{Bambah}
\begin{eqnarray}
\hat{{\cal N}}_{+} = \hat{N}_{+} F(\hat{C},\hat{N}_0),
\label{lad13}
\end{eqnarray}
where $\hat{{\cal N}}_{+}$ is the new deformed ladder operator and
$\displaystyle{F(\hat{C}, \hat{N}_0) = \frac{\hat{N}_0 + \delta}{\hat{C}-h(\hat{N}_0)}=\frac{\hat{N}_0 + \delta}{\hat{N}_{-}\hat{N}_{+}}}$, with $\delta$ is a parameter.

We can generate Heisenberg algebra, for the system (\ref{se}),
through the newly deformed ladder operator (\ref{lad13}) in the form \cite{chi3}
\begin{eqnarray}
\fl \mbox{Case: (i)}\;\; [\hat{N}_{-}, \hat{{\cal N}}_{+}]|n\rangle = |n\rangle, \;\; [\hat{\cal N}_{+} \hat{N}_{-}, \hat{N}_{-}]|n\rangle = - \hat{N}_{-}|n\rangle, \;\;
[\hat{\cal N}_{+} \hat{N}_{-}, \hat{\cal N}_{+}]|n\rangle = \hat{\cal N}_{+}|n\rangle.
\label{lad14}
\end{eqnarray}

Similarly by rescaling the ladder operator $\hat{N}_{-}$ such a way that
\begin{eqnarray}
\hat{\cal N}_{-} = F(\hat{C}, \hat{N}_0) \hat{N}_{-},
\label{lad13b}
\end{eqnarray}
where $\displaystyle{F(\hat{C}, \hat{N}_0) = \frac{\hat{N}_0 + \delta}{\hat{C}-h(\hat{N}_0)}=\frac{\hat{N}_0 + \delta}{\hat{N}_{-}\hat{N}_{+}}}$, we can generate the second set of Heisenberg algebra in the form
\begin{eqnarray}
\fl \mbox{Case: (ii)}\;\;  [\hat{\cal N}_{-}, \hat{N}_{+}]|n\rangle = |n\rangle, \;\; [\hat{N}_{+}\hat{\cal N}_{-}, \hat{\cal N}_{-}]|n\rangle = -\hat{\cal N}_{-}|n\rangle, \;\;
[\hat{N}_{+}\hat{\cal N}_{-}, \hat{N}_{+}]|n\rangle = \hat{N}_{+}|n\rangle.
\label{lad15}
\end{eqnarray}

The constant $\delta$ in $F(\hat{C}, \hat{N}_0)$ can be fixed by utilizing the commutation relations,
$[\hat{N}_{-}, \hat{\cal N}_{+}]|3\rangle = |3\rangle$ and $[\hat{\cal N}_{-}, \hat{N}_{+}]|3\rangle = |3\rangle$.
From these two relations, we find $\delta = -2$  and fix
${\displaystyle F(\hat{C}, \hat{N}_0) = \frac{\hat{N}_0 - 2}{\hat{N}_{-} \hat{N}_{+}}}$.

Finally, one can rescale both the operators $\hat{N}_{+}$ and $\hat{N}_{-}$ simultaneously and evaluate the
commutation relations. For example, let us rescale $\hat{N}_{+}$ and $\hat{N}_{-}$
respectively as $\hat{K}_{+} = \hat{N}_{+}G(\hat{C}, \hat{N}_0)$ and
$\hat{K}_{-} = G(\hat{C}, \hat{N}_0) \hat{N}_{-}$. The explicit form of $G(\hat{C}, \hat{N}_0)$ can then be found
by using the commutation relation $[\hat{K}_{-}, \hat{K}_{+}] = \hat{I}$, that is
\begin{eqnarray}
G(\hat{C}, \hat{N}_0) \hat{N}_{-} \hat{N}_{+} G(\hat{C}, \hat{N}_0) - \hat{N}_{+} G^{2}(\hat{C}, \hat{N}_0) \hat{N}_{-} = \hat{I}.
\label{case3}
\end{eqnarray}
Solving (\ref{case3}) we find $G(\hat{C}, \hat{N}_0) = \sqrt{F(\hat{C}, \hat{N}_0)}$.

With this choice of $G(\hat{C}, \hat{N}_0)$ we can establish
\begin{eqnarray}
\fl \mbox{Case: (iii)}\; [\hat{K}_{-}, \hat{K}_{+}]|n\rangle = |n\rangle, \;\; [\hat{K}_{0}, \hat{K}_{-}]|n\rangle = -\hat{K}_{-}|n\rangle,  \;\;\;
[\hat{K}_{0}, \hat{K}_{+}]|n\rangle = \hat{K}_{+}|n\rangle,
\label{lad16}
\end{eqnarray}
where $\hat{K}_{0} = \hat{K}_{+} \hat{K}_{-}$. Here $\hat{K}_0$ serves as a number operator.

We construct squeezed and nonlinear squeezed states using these three sets of new deformed ladder operators.

\section{Nonlinear squeezed states}
\subsection{Non-unitary squeezing operators and nonlinear squeezed states}
The transformed operators $\hat{\cal N}_{+}$ and $\hat{\cal N}_{-}$ which satisfy the commutation relations
(\ref{lad14}) and (\ref{lad15}) help us to define two non-unitary squeezing operators, namely
\begin{eqnarray}
\mbox{Case: (i)}\;\;\;\;S(\beta) &=& e^{\frac{\beta}{2} \hat{\cal N}^{2}_{+} - \frac{\beta^{*}}{2} \hat{N}^2_{-}},\\
\mbox{Case: (ii)}\;\;\;\tilde{S}(\beta) &=& e^{\frac{\beta}{2} \hat{N}^{2}_{+}- \frac{\beta^{*}}{2} \hat{\cal N}^2_{-}}.
\label{lad25a}
\end{eqnarray}

By applying these operators on the lowest energy state $|3\rangle$ given in (\ref{eigfn}), we
obtain the nonlinear squeezed states as
\begin{eqnarray}
\fl \;\mbox{Case: (i)}\;\;\;\;\;|\beta, \tilde{f}\rangle &=& N_{\beta}\sum^{\infty}_{n=0}\frac{\beta^n}{2^n \; n!}\sqrt{\frac{(2n)!}{(2n +2)!\;(2n+3)!}}\;|2n+3\rangle,
\label{lad26b}\\
\fl \;\mbox{Case: (ii)}\;\;\;\;\widetilde{|\beta, \tilde{f}\rangle} &=& {\tilde{N}_{\beta}}
\sum^{\infty}_{n=0}\frac{\beta^n}{2^n\;n!} \sqrt{(2n)!(2n+2)!(2n+3)!}\;|2n+3\rangle,
\label{lad27b}
\end{eqnarray}
where the normalization constant $N_{\beta}$ and $\tilde{N}_{\beta}$ are given by
\begin{eqnarray}
\fl \;\;\;\;\mbox{Case: (i)}\;\;\;\;N_{\beta} = \left(\sum^{\infty}_{n=0}\frac{|\beta|^{2n} (2n)! }{4^{ n} (n!)^2 (2n+2)!(2n+3)!}\right)^{-1/2},
\\
\fl \;\;\;\;\mbox{Case: (ii)}\;\;\;\tilde{N}_{\beta} = \left(\sum^{\infty}_{n=0}{\frac{|\beta|^{2n}\;(2n)!\;(2n+2)!\;(2n+3)!}{4^{ n} 
(n!)^2}}\right)^{-1/2}.
\label{norm2}
\end{eqnarray}
The series given in (\ref{norm2}) is of the form ${\displaystyle \sum^{\infty}_{n=0} \frac{12 |\beta|^{2 n}}{[x_n]!}}$, 
with ${\displaystyle x_{n} = \frac{2 n}{(2 n -1)(2n+1)(2n+2)^2(2n+3)}}$ and $[x_n]! = x_n . x_{n-1}. . . x_1$. 
One can unambiguously prove that the series given in (\ref{norm2}) is a divergent one  
since for non-zero values of $|\beta|$, the limit yields ${\displaystyle L^2 = \lim_{n\to \infty} x_n = 0}$  
and consequently it does not meet the necessary condition, $|\beta| < L$ with $L^{2} \neq 0$.  
Since $\tilde{N}_{\beta} = 0$, the dual states (\ref{lad27b}) do not exist. Hence, we conclude
that for the generalized isotonic oscillator one can construct only nonlinear squeezed
states and not their dual counterpart.

\subsection{Unitary squeezing operator and squeezed states}
In the Case (iii) the squeezing operator
\begin{eqnarray}
S(\xi) = e^{\frac{\xi}{2} \hat{K}^{2}_{+} - \frac{\xi^*}{2} \hat{K}^2_{-}}
\label{lad28}
\end{eqnarray}
is an unitary one.
By applying this operator on the lowest energy state $|3\rangle$
given in (\ref{eigfn}), we get the normalized form of squeezed states as
\begin{eqnarray}
\fl \;\mbox{Case: (iii)}\;\;\;\;|\xi \rangle =  S(\xi) |3\rangle = N_{\xi}\sum^{\infty}_{n=0}\frac{\xi^{n}}{2^n n!}\sqrt{(2n)!}|2n+3\rangle,
\label{lad29}
\end{eqnarray}
where $N_{\xi}$ can be obtained from the normalization condition $\langle \xi|\xi\rangle = 1$. Doing so we find the 
normalization constant 
\begin{eqnarray}
\fl \quad \qquad \qquad \qquad \quad N_{\xi} = \left(\sum^{\infty}_{n=0}\frac{|\xi|^{2n}\;(2n)!}{4^{n}(n!)^2}\right)^{-1/2}.
\label{lad30}
\end{eqnarray}

These squeezed states $|\xi\rangle$ are in the same form as that of harmonic oscillator \cite{walls}.
We will discuss the properties of these states separately hereafter.

\section{Non-classical properties}
In this section we study certain photon statistical properties, namely (i) the photon number distribution $(P(n))$,
(ii) Mandel's parameter $(Q)$ and (iii) the second-order correlation function $(g^2 (0))$ associated with the
nonlinear squeezed states given in (\ref{lad26b}) and squeezed states given in (\ref{lad29}).
In addition to these, we also analyze
quadrature and amplitude-squared squeezing for the non-classical states.

\subsection{Photon statistical properties}
To start with, we calculate the probability of finding $n$ photons in the nonlinear
squeezed states (\ref{lad26b}) which in turn gives
\begin{eqnarray}
\fl \;\;\;\;\mbox{Case: (i)}\;\;\;P(2n) = |\langle 2n+3 | \beta, \tilde{f}\rangle|^2 
                                        = \frac{N^2_{\beta}|\beta|^{2 n} (2n)!}{4^{n} (n!)^2\;(2n+2)!\;(2n+3)}. 
\label{ps1}
\end{eqnarray}

\begin{figure}[!ht]
\vspace{0.1cm}
\begin{center}
\includegraphics[width=1\linewidth]{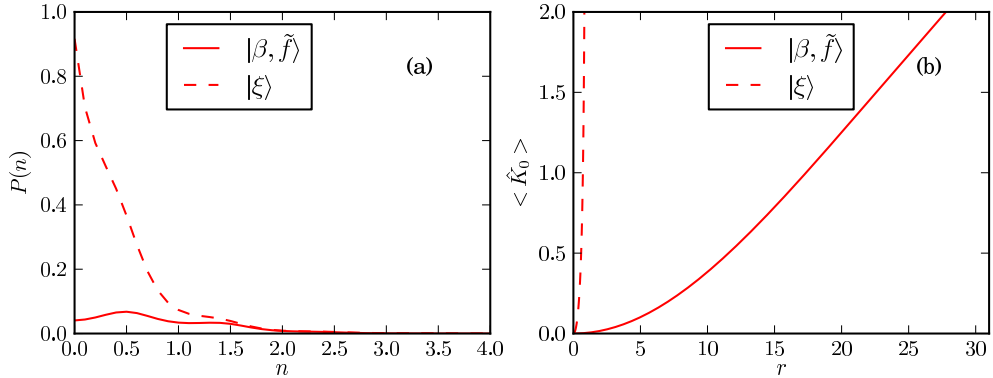}
\end{center}
\vspace{-6cm}
\caption{The plots of (a) photon number distribution  $P(n)$ and
(b) average number of photons $\langle \hat{K}_{0} \rangle$ in
nonlinear squeezed states (\ref{lad26b})
and squeezed states (\ref{lad29})}.
\label{ss_phot}
\vspace{-0.3cm}
\end{figure}

The photon number distribution for the nonlinear squeezed states
$|\beta, \tilde{f}\rangle$ is calculated ($r = |\beta| = 20$ with $n_{max} = 70$) 
and plotted in figure \ref{ss_phot}(a). The result 
confirms that the distribution is not a Poissonian one.

Since $\hat{K}_{+}, \hat{K}_{-}$ and $\hat{K}_{0}$ act on the states
$|3\rangle, |4\rangle, |5\rangle,...$ in the same
way as creation $(\hat{a}^{\dagger})$, annihilation $(\hat{a})$ and
number $(\hat{n})$ operators act on the states
$|0\rangle, |1\rangle, |2\rangle,...$  of harmonic oscillator potential,
we consider $\hat{K}_{0}$ as number operator for
the system (\ref{se}) in the sub-Hilbert space spanned by the eigenstates
$|3\rangle, |4\rangle, |5\rangle, ...$.
So, we examine Mandel's parameter $Q$ and second-order correlation function $g^{2}(0)$ in
terms of $\hat{K}_{0}$ only \cite{Mandel, Paul, Mah, Ant}, that is 
\begin{eqnarray}
Q = \frac{\langle \hat{K}^2_{0} \rangle}{\langle\hat{K}_0\rangle}-\langle\hat{K}_0\rangle-1, \qquad \;\;\;
g^{(2)}(0) = \frac{\langle \hat{K}^2_{0} \rangle-\langle\hat{K_0}\rangle}{\langle \hat{K}_{0} \rangle^2}.
\label{g}
\end{eqnarray}

\begin{figure}[!ht]
\vspace{-0.2cm}
\begin{center}
\includegraphics[width=1\linewidth]{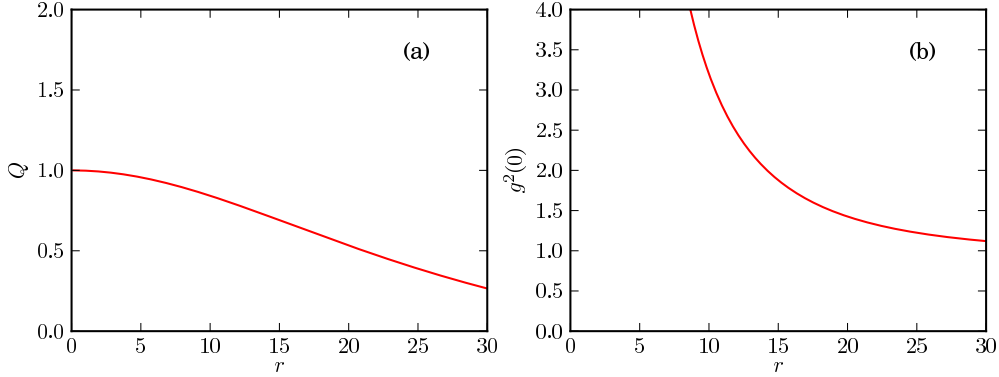}
\end{center}
\vspace{-5.9cm}
\caption{The plots of (a) Mandel's parameter $Q$ and (b) the second-order correlation function $g^{2}(0)$
of the nonlinear squeezed states (\ref{lad26b}).} \label{ss_prop}
\vspace{-0.3cm}
\end{figure}

To calculate Mandel's parameter, we first obtain expressions for
$\langle \hat{K}_{0} \rangle$ and $\langle \hat{K}^{2}_0 \rangle$
corresponding to the nonlinear squeezed states given in (\ref{lad26b}),
which are of the form
\begin{eqnarray}
\fl \mbox{Case: (i)}\;\;\;\langle \hat{K}_0 \rangle &=& N^2_{\beta} \sum^{\infty}_{n=1}
{\frac{|\beta|^{2 n}\;(2n-1)!}{4^{n-1}\;((n-1)!)^2\;(2n+2)!\;(2n+3)!}},
\label{c1_ss_N1}\\
\fl \qquad \qquad \;\langle \hat{K}^{2}_0 \rangle &=& N^2_{\beta} \sum^{\infty}_{n=1}
{\frac{|\beta|^{2n}\;(2n)!}{4^{n-1}\;((n-1)!)^2\;(2n+2)!\;(2n+3)!}},
\label{c1_ss_N2}
\end{eqnarray}
where $\langle \hat{K}_{0} \rangle$ gives the average number of photons in the nonlinear
squeezed states $|\beta, \tilde{f}\rangle$
for different values of $r$. The results are plotted in figure \ref{ss_phot}(b) which
demonstrates the nonlinear dependency between $\langle \hat{K}_{0} \rangle$ and $r$.

\begin{figure}[!ht]
\vspace{-0.2cm}
\begin{center}
\includegraphics[width=1\linewidth]{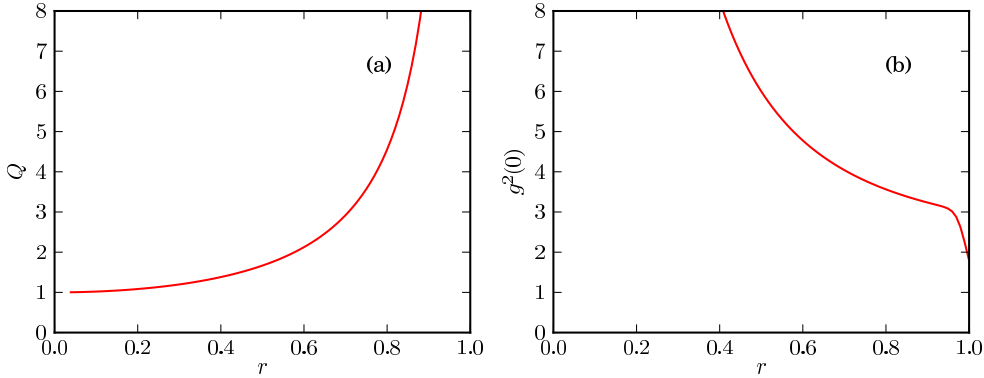}
\end{center}
\vspace{-5.9cm}
\caption{The plots of (a) Mandel's parameter $Q$ and (b) the second-order correlation function $g^{2}(0)$
of the squeezed states (\ref{lad29}).} \label{ss_prop_c3}
\vspace{-0.3cm}
\end{figure}

Substituting the expressions (\ref{c1_ss_N1})-(\ref{c1_ss_N2}) in  (\ref{g}) and evaluating
the resultant expressions we can obtain the Mandel's parameter and second-order correlation 
function for the states $|\beta, \tilde{f}\rangle$. Here we investigate the variations of $Q$ and $g^{2}(0)$
against $r ( < 31)$ and summarize the results in figures \ref{ss_prop}(a)
and \ref{ss_prop}(b). From the figures we observe that
for the values of $r (< 31)$ with $n_{max} = 70$,  $Q > 0$ and $g^{2}(0) > 1$. The positive values of 
$g^{2}(0)$ indicate the super-Poissonian
nature of the nonlinear squeezed states $|\beta, \tilde{f}\rangle$.

The photon number distribution for the states (\ref{lad29}) corresponding to the Case (iii) are found to be
\begin{eqnarray}
\fl \qquad \quad \mbox{Case:\;(iii)}\;\;\;P(2n) = |\langle 2 n+ 3{|\xi, \tilde{f}\rangle}|^2 = N^2_{\xi} \frac{|\xi|^{2 n} (2n)!} {4^n\;(n!)^2},
\label{ps3}
\end{eqnarray}
which is calculated and plotted in figure \ref{ss_phot}(a) with $r = 0.4$ and $n_{max} = 70$. As shown in the
figure, the photon number distribution for the states $|\xi\rangle$ is not a
Poissonian one.

The Mandel's parameter and second-order correlation function
for the squeezed states $|\xi\rangle$ are found to be
\begin{eqnarray}
\fl \quad \quad \;\langle \hat{K}_0 \rangle =  N^2_{\xi} \sum^{\infty}_{n=1} |\xi|^{2n} \frac{(2n-1)!}{4^{n-1}\;((n-1)!)^2},
\quad \quad 
\langle \hat{K}^2_0 \rangle =  N^2_{\xi} \sum^{\infty}_{n=1} |\xi|^{2n} \frac{(2n)!}{4^{n-1}\;((n-1)!)^2},
\label{expss3}
\end{eqnarray}
where $\langle \hat{K}_0 \rangle$ is the average value of number of
photons in the squeezed states which is plotted in figure \ref{ss_phot}(b).
Substituting (\ref{expss3}) in (\ref{g}), we can calculate Mandel's parameter ($Q$)
and the second-order correlation function ($g^{2}(0)$)
for the squeezed states given in (\ref{lad29}). In figures \ref{ss_prop_c3}(a) and \ref{ss_prop_c3}(b), 
the parameters $Q$ and $g^{2}(0)$ of the states $|\xi\rangle$ are shown as a function of $r$. The states 
given in equation (\ref{lad29}) exhibit super-Poissonian feature for a range of $r$. 

\subsection{$A_3$-parameter}
In addition to the above non-classical properties, one can also investigate the parameter $A_3$ which
was introduced by Agarwal and Tara \cite{agarwal1}. It was also recently studied for the newly introduced
$\beta$- nonlinear coherent states \cite{tavassoly_beta}. The parameter $A_3$ can be calculated from the
expression \cite{agarwal1},

\begin{eqnarray}
A_{3} = \frac{det\; m^{(3)}}{det\; \mu^{(3)} - det\; m^{(3)}},
\label{A3}
\end{eqnarray}
where
$     m^{(3)}= \left(\begin{array}{ccc}
    1 & m_1 & m_2 \\
    m_1 & m_2 & m_3 \\
    m_2& m_3 & m_4\\
    \end{array} \right)$
and
    $ \mu^{(3)}= \left(
  \begin{array}{ccc}
    1 & \mu_1 & \mu_2 \\
    \mu_1 & \mu_2 & \mu_3 \\
    \mu_2& \mu_3 & \mu_4\\
    \end{array}
    \right).$
\\
\\

In the above, $m_j= \hat{K}^{j}_{+} \hat{K}^j_{-} $ and $\mu_j=(\hat{K}_{+} \hat{K}_{-})^j$,\;$j = 1, 2, 3, 4 $.
For the coherent and vacuum states $\det \;m^{(3)}=0$ and for a Fock state $\det \;m^{(3)}= -1$ and $\det \;\mu^{3} = 0$.
For the non-classical states $\det \;m^{(3)} < 0$ and since $\det \;\mu^{(3)} > 0$, it follows that parameter $A_3$
lies between 0 and -1.

\vspace{0.2cm}
\begin{figure}[ht]
\hspace{3.1cm}\includegraphics[width=1\linewidth]{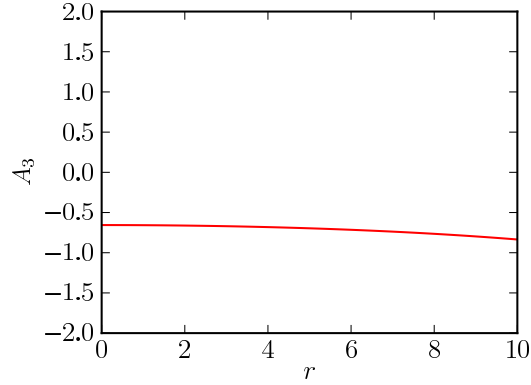}
\vspace{-5.5cm}
\caption{The plot of parameter $A_3$.}
\label{a3_nlss}
\end{figure}

To obtain an expression for  parameter $A_3$, one has to evaluate $\langle m_j \rangle$'s and
$\langle \mu_j \rangle$'s, $j=1,2,3,4$, with respect to the nonlinear squeezed states $|\beta, \tilde{f}\rangle$.
 Let us first calculate $\langle m_j \rangle$:
\begin{eqnarray}
\fl \qquad \qquad m_j |\beta, \tilde{f}\rangle =  N_{\beta} \sum^{\infty}_{n=0}\frac{\beta^{n}}{2^{n} n!}\sqrt{\frac{(2n)!}{(2n+2)!\;(2n+3)!}}\;m_j|2n+3\rangle.
\label{sm1}
\end{eqnarray}
Since $m_j |2n+3\rangle = K^{j}_{+} K^{j}_{-}|2n+3\rangle = 2n (2n-1) (2n-2)...(2n-j+1) |2n+3\rangle$, we get
\begin{eqnarray}
\fl \qquad \qquad m_j |\beta, \tilde{f}\rangle =  N_{\beta} \sum^{\infty}_{n=0}\frac{\beta^{n}}{2^n\;n!}\sqrt{\frac{(2n)!}{\;(2n+2)!\;(2n+3)!}}\nonumber \\
\qquad \qquad \qquad \qquad \times\;2n (2n-1)...(2n-j+1)\;|2n+3\rangle.
\label{sm2a}
\end{eqnarray}
Using (\ref{sm2a}), we find
\begin{eqnarray}
\fl \qquad \qquad  \langle \beta, \tilde{f}|m_j|\beta, \tilde{f}\rangle =  N^2_{\beta} \sum^{\infty}_{n= \lceil \frac{j}{2} \rceil}\frac{|\beta|^{2n}\;((2n)!)^2}{(2n-j)! 4^{n}(n!)^2\;(2n+2)!\;(2n+3)!}, 
\label{sm2}
\end{eqnarray}
where $\lceil \frac{j}{2} \rceil$ is  ceiling($\frac{j}{2}$). Using these expressions, we calculate parameter $A_{3}$ for the
nonlinear squeezed states $|\beta, \tilde{f}\rangle$. The result is given in
figure \ref{a3_nlss}. The figure confirms that the 
value of parameter  $A_3$ lies in between $-1$ and $0$ for all positive values of $r$. The negative 
values prove the non-classical nature of the nonlinear squeezed states.

\subsection{Quadrature squeezing}
To study the non-classical nature of the squeezed states, we define two
Hermitian operators, namely $\hat{x}$ and $\hat{p}$ in terms of the deformed creation and annihilation operators,
$\hat{K}_{+}$ and $\hat{K}_{-}$ in the form \cite{walls, yuen,  squeezing}
\begin{eqnarray}
\hat{x} = \frac{1}{\sqrt{2}}(\hat{K}_{+} + \hat{K}_{-}), \qquad \hat{p} = \frac{i}{\sqrt{2}}(\hat{K}_{+} - \hat{K}_{-}).
\label{h1}
\end{eqnarray}
The operators $\hat{x}$ and $\hat{p}$ satisfy the commutation relation $[\hat{x}, \hat{p}] = i$.

\vspace{0.2cm}
\begin{figure}[ht]
\centering
\includegraphics[width=0.85\linewidth]{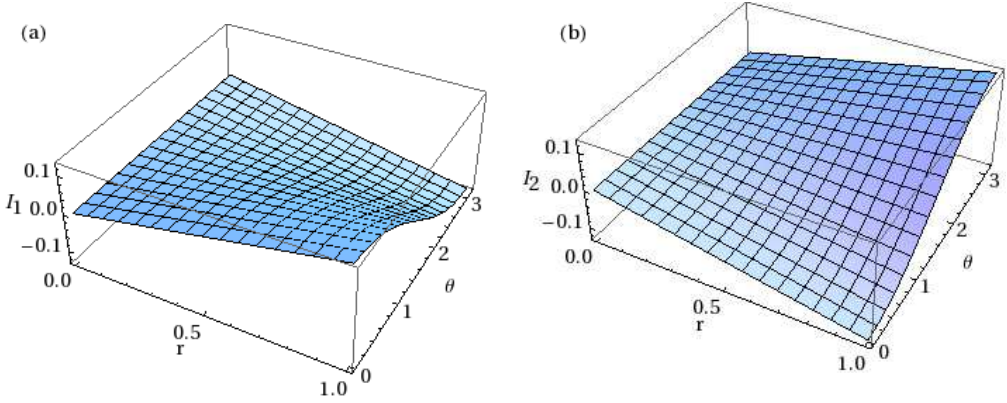}
\vspace{-0.1cm}
\caption{The plots of the identities (a)  $I_{1}$ and  (b) $I_{2}$ calculated with
respect to nonlinear squeezed states (\ref{lad26b}) with $n_{max} = 70$.}
\label{c1_ss_id12}
\vspace{-0.3cm}
\end{figure}

The squeezed states (\ref{lad26b}) and (\ref{lad29}) satisfy the Heisenberg
uncertainty relation $(\Delta \hat{x})^2 (\Delta \hat{p})^2 \geq \frac{1}{4}$ .
A state is said to be squeezed in $\hat{x}$ or $\hat{p}$,  if
$(\Delta \hat{x})^2 < \frac{1}{2}$ or $(\Delta \hat{p})^2 < \frac{1}{2}$.
Here, $\Delta \hat{x}$ and $\Delta \hat{p}$ denote the uncertainties in $\hat{x}$ and $\hat{p}$ respectively.
The squeezing conditions can be transformed to the forms  \cite{tavanon}
\begin{eqnarray}
 \fl \qquad I_{1} &=& \langle \hat{K}^2_{-} \rangle + \langle \hat{K}^2_{+} \rangle - \langle \hat{K}_{-} \rangle^{2} - \langle \hat{K}_{+} \rangle^{2}
- 2 \langle \hat{K}_{-} \rangle \langle \hat{K}_{+} \rangle + 2\langle \hat{K}_{+} \hat{K}_{-} \rangle < 0,
\label{id1}\\
 \fl \qquad I_{2} &=& -\langle \hat{K}^2_{-} \rangle - \langle \hat{K}^2_{+} \rangle + \langle \hat{K}_{-} \rangle^{2} + \langle \hat{K}_{+} \rangle^{2}
- 2 \langle \hat{K}_{-} \rangle \langle \hat{K}_{+} \rangle + 2\langle \hat{K}_{+} \hat{K}_{-} \rangle < 0,
\label{id2}
\end{eqnarray}
where the expectation values are to be calculated with respect to squeezed states for which the quadrature squeezing
has to be examined.

The identities, (\ref{id1}) and (\ref{id2}), are calculated for the nonlinear squeezed states
(\ref{lad26b}) and presented in figures \ref{c1_ss_id12}(a) and \ref{c1_ss_id12}(b) respectively
with $\beta = r e^{i\;\theta}$.

\begin{figure}[ht]
\centering
\includegraphics[width=0.85\linewidth]{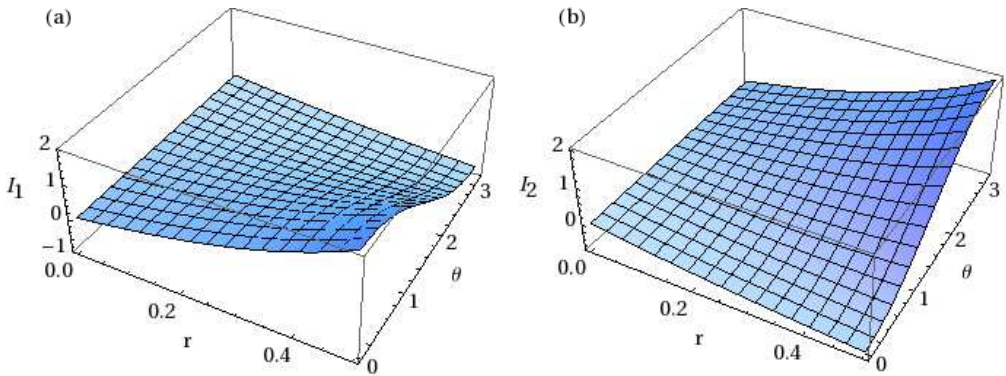}
\vspace{-0.1cm}
\caption{The plots of the identities (a) $I_{1}$ and  (b) $I_{2}$ calculated with
respect to squeezed states (\ref{lad29}) with $n_{max} =  70$.}
\label{c3_ss_id12}
\vspace{-0.3cm}
\end{figure}

From figures \ref{c1_ss_id12}(a) and \ref{c1_ss_id12}(b), we observe that the identities
(\ref{id1}) and (\ref{id2}) for the nonlinear squeezed states $|\beta, \tilde{f}\rangle$
satisfying the uncertainty relation
show small oscillations in $I_1$ and $I_2$. These two quantities, $I_1$ and $I_2$, oscillate out of phase
$\pi$ with each other. In other words the squeezing can be observed in both
the quadratures, $\hat{x}$ and $\hat{p}$, at different values of $\theta$.

The same type of squeezing is observed in the squeezed states (\ref{lad29}) as well, which is depicted in
figures \ref{c3_ss_id12}(a) and \ref{c3_ss_id12}(b). The squeezing shown by the nonlinear squeezed states (\ref{lad26b})
and squeezed states (\ref{lad29}) confirm the non-classical nature of the
associated states.

\subsection{Amplitude-squared squeezing}
The amplitude-squared squeezing, which was introduced by Hillery \cite{hillery}, involves two operators
which represent the real and imaginary parts of the square of the amplitude of a radiation field.
To investigate the amplitude-squared squeezing effect, we introduce again two Hermitian operators $\hat{X}$ and $\hat{P}$
from $\hat{K}_{+}$ and $\hat{K}_{-}$ respectively of the form
\begin{eqnarray}
\hat{X} = \frac{1}{\sqrt{2}}(\hat{K}^2_{+} + \hat{K}^2_{-}), \qquad
\hat{P} = \frac{i}{\sqrt{2}}(\hat{K}^2_{+}- \hat{K}^2_{-}).
\label{as3}
\end{eqnarray}

Here $\hat{X}$ and $\hat{P}$ are the operators corresponding to the real and imaginary parts
of the square of the complex amplitude of a radiation field. The Heisenberg uncertainty relation
of these conjugate operators is then given by
$(\Delta \hat{X})^2 (\Delta \hat{P})^2 \ge -\frac{1}{4}\langle[\hat{X}, \hat{P}]\rangle^2$.
For the nonlinear squeezed states (\ref{lad26b}) and the squeezed states (\ref{lad29}), we find
$(\Delta \hat{X})^2 < -\frac{i}{2} \langle[\hat{X}, \hat{P}]\rangle$ or
$(\Delta \hat{P})^2 < -\frac{i}{2} \langle[\hat{X}, \hat{P}]\rangle$
which in turn confirm that the states are non-classical.
The conditions for the amplitude-squared squeezing read \cite{tavanon}
\begin{eqnarray}
\fl \qquad I_{3}  = \frac{1}{4} \left(\langle {\hat{K}_{-}}^4 \rangle + \langle {\hat{K}_{+}}^4\rangle
                     - \langle {\hat{K}_{-}}^2 \rangle^{2} - \langle {\hat{K}_{+}}^2 \rangle^{2}
                     - 2 \langle {\hat{K}_{-}}^2 \rangle \langle {\hat{K}_{+}}^2 \rangle
                      +\langle {\hat{K}_{+}}^2 {\hat{K}_{-}}^2 \rangle + \langle {\hat{K}_{-}}^2 {\hat{K}_{+}}^2 \rangle \right) \nonumber
         \\ - \langle {\hat{K}_{+}} {\hat{K}_{-}} \rangle - \frac{1}{2} < 0,
\label{id3}\\
\fl \qquad I_{4} = \frac{1}{4} \left(-\langle {\hat{K}_{-}}^4 \rangle - \langle {\hat{K}_{+}}^4\rangle
                   +\langle {\hat{K}_{-}}^2 \rangle^{2}+ \langle {\hat{K}_{+}}^2 \rangle^{2}
                   - 2 \langle {\hat{K}_{-}}^2 \rangle \langle {\hat{K}_{+}}^2 \rangle
                      +\langle {\hat{K}_{+}}^2 {\hat{K}_{-}}^2 \rangle + \langle {\hat{K}_{-}}^2 {\hat{K}_{+}}^2 \rangle \right) \nonumber
         \\ - \langle {\hat{K}_{+}} {\hat{K}_{-}} \rangle - \frac{1}{2} < 0,
\label{id4}
\end{eqnarray}
where the expectation values are to be calculated with respect to the nonlinear squeezed states
$|\beta, \tilde{f}\rangle$ for which the
amplitude-squared squeezing property has to be examined.

\begin{figure}[ht]
\centering
\includegraphics[width=0.95\linewidth]{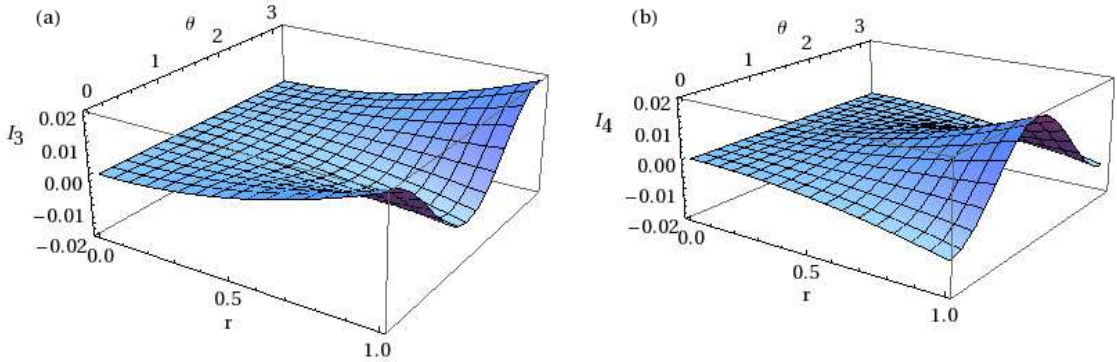}
\vspace{-0.1cm}
\caption{The plots of the identities (a)  $I_{3}$,  and (b) $I_{4}$ calculated with
respect to nonlinear squeezed states (\ref{lad26b}) for $n_{max} = 70$.}
\label{c1_ss_id34}
\vspace{-0.3cm}
\end{figure}

We evaluate the identities, (\ref{id3}) and (\ref{id4}), numerically and
plot the results in figures \ref{c1_ss_id34}(a) and \ref{c1_ss_id34}(b).
The identities $I_3$ and $I_4$ also vary in an oscillatory manner.
For certain values of $r$ and $\theta$ when one of the identities $I_3\;($or $I_4)$ is positive the
other identity $I_4\;($or $I_3)$ becomes negative. The negativity of $I_3\;(I_4)$ indicates the amplitude squared squeezing in 
 $\hat{X}\;(\hat{P})$ operators respectively.

\vspace{0.2cm}
\begin{figure}[ht]
\centering
\includegraphics[width=0.95\linewidth]{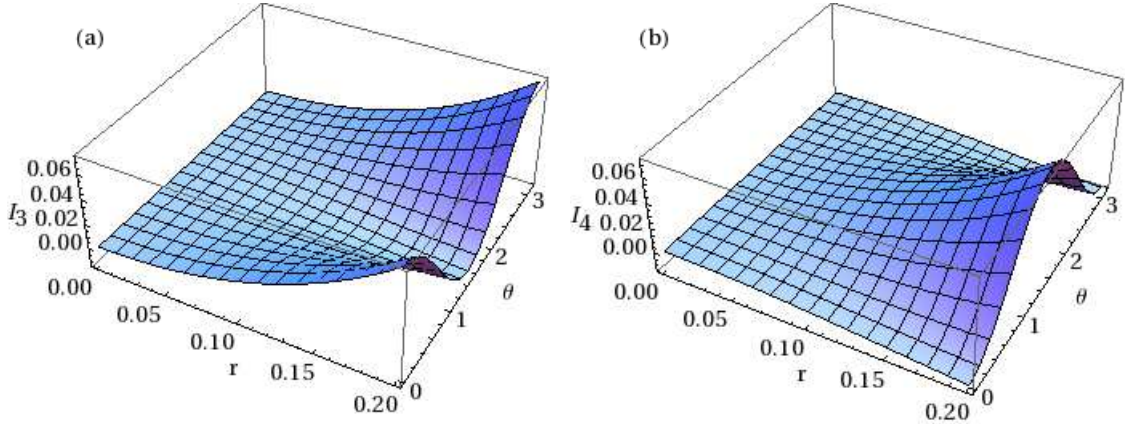}
\vspace{-0.1cm}
\caption{The plots of the identities (a) $I_{3}$, and (b) $I_{4}$ calculated with
respect to squeezed states  (\ref{lad29}).}
\label{c3_ss_id34}
\vspace{-0.3cm}
\end{figure}

We calculate the identities $I_3$ and $I_4$ in (\ref{id3}) and (\ref{id4})
for the squeezed states (\ref{lad29}) and plot the results in figure \ref{c3_ss_id34}(a) and \ref{c3_ss_id34}(b),
which in turn confirm the non-classical nature of the
squeezed states.

\section{Quadrature distribution and quasi-probability distribution functions}

\subsection{Phase-parameterized field strength distribution}
We study phase-parameterized distribution for the nonlinear squeezed states $|\beta, \tilde{f}\rangle$,
in order to analyze the nature of the dependency of quantum noise on phase, which is defined to be \cite{obada}
\begin{eqnarray}
P(x, \phi) = |\langle x, \phi{|\beta, \tilde{f}\rangle}|^{2},
\label{para2}
\end{eqnarray}
where $|x,\phi\rangle$ is the eigenstate of the quadrature component
$\hat{x}(\phi) = \frac{1}{\sqrt{2}}\left(e^{- i \phi} \hat{K}_{-} + e^{i \phi} \hat{K}_{+} \right)$.
In other words
\begin{equation}
\hat{x}(\phi)|x,\phi\rangle = x|x,\phi\rangle,
\label{xeq}
\end{equation}
which can be expressed in photon number basis in the form
\begin{eqnarray}
|x,\phi\rangle = \frac{e^{\frac{-x^2}{2}}}{\pi^{\frac{1}{4}}} \sum^{\infty}_{n=0} \frac{H_n(x) e^{i n \phi }}{\sqrt{2^n n!}}|n+3\rangle,
\label{xphi}
\end{eqnarray}
where $H_{n}(x)$ is the Hermite polynomial. Substituting (\ref{xphi}) in (\ref{para2}) with $\beta = r e^{i \theta}$ we obtain
\begin{eqnarray}
\fl \quad P(x, \phi) =\frac{{N}^{2}_{\beta} e^{-x^2}}{\sqrt{\pi}} \sum^{\infty}_{n,m = 0} \left(\frac{r}{4}\right)^{n+m}\frac{
H_{2n}(x)H_{2m}(x)}{n!\;m!} \nonumber \\
\quad \qquad \quad \times\frac{\cos[(m-n)(2\phi - \theta)]}{\sqrt{(2n+2)!\;(2m+2)!\;(2n+3)!\;(2m+3)!}}.
\label{pa_nlss_1}
\end{eqnarray}
\begin{figure}[!ht]
\vspace{-0.5cm}
\begin{center}
\includegraphics[width=0.5\linewidth]{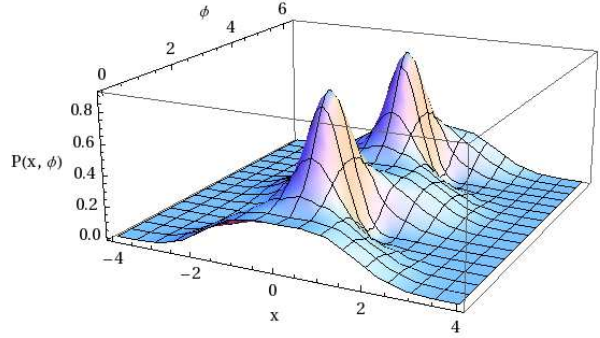}
\end{center}
\vspace{-0.5cm}
\caption{The plot of $P(x,\phi)$ which is calculated with respect to
squeezed states (\ref{lad26b})}
\vspace{-0.3cm}
\label{ss_phase}
\end{figure}

From the expressions (\ref{pa_nlss_1}), we determine the quadrature function numerically
with $\beta = r e^{i \theta}$. The numerical results are displayed in figure
\ref{ss_phase} with $r = 10$ and $\theta = 0.5$ for the nonlinear
squeezed states $|\beta, \tilde{f}\rangle$.  The figure \ref{ss_phase} shows an oscillating wave packet around  $x = 0$ with two peaks near $\phi =  \frac{\pi}{2}$ and $\frac{3 \pi}{2}$. When $|x| > 3$, the phase information
$P(x,\;\phi)$ disappears. The quadrature distribution $P(x,\; \phi)$ plotted in figure \ref{ss_phase} depicts the time evolution of 
position probability density of the squeezed vacuum state during one oscillation period. In fact, this quadrature distribution plot matches with the experimental result  
reported in Ref. \cite{nat}. 

\subsection{$s$-parameterized quasi-probability function}
In this sub-section, we study $s$-parameterized quasi-probability distribution function for the
nonlinear squeezed states (\ref{lad26b}). The $s$-parameterized quasi-probability
distribution function is defined as the Fourier transform of the $s$-parameterized characteristic
function \cite{spara, vbook,barn}
\begin{eqnarray}
F(z, s) = \frac{1}{\pi^2}\int C(\lambda, s) e^{(\lambda^* z - \lambda z^*)} d^{2} \lambda,
\label{spara}
\end{eqnarray}
where
\begin{eqnarray}
C(\lambda, s) = \mbox{Tr}[\hat{\rho} D(\lambda)] \exp{\left[\frac{s}{2} |\lambda|^2\right]}
\label{char}
\end{eqnarray}
is the $s$-parameterized characteristic function \cite{obada} and $D(\lambda)$ is the
displacement operator. To study the quasi-probability distribution for
the nonlinear squeezed states constructed for the system (\ref{se}),
we consider the unitary displacement operator $D(\lambda) = \exp{(\lambda \hat{K}_{+} - \lambda^{*} \hat{K}_{-})}$ from Case (iii)
since $\hat{K}_{-}$ and $\hat{K}_{+}$  act as annihilation and
creation operators $\hat{a}$ and $\hat{a}^{\dagger}$.
This $s$-parameterized function is introduced by Cachill and Glauber with $s$ being a continuous variable \cite{spara}.
This function is
known as the generalized function that interpolates the Glauber-Sudarhsan $P$-function for $s$ = 1, Wigner function $W$ for $s = 0$
and Husimi $Q$-function for $s = -1$
\cite{spara}. The quasi-probability distribution functions provides insight into
the non-classical features of the radiation field.

The characteristic function (\ref{char}) for the squeezed states read
\begin{eqnarray}
C(\lambda, s) = \exp{\left[\frac{s}{2} |\lambda|^2\right]} \sum^{\infty}_{m,\;n=0}
{B}_{n,\;m}\langle 2m+3|D(\lambda)|2n+3\rangle,
\label{chars}
\end{eqnarray}
where the coefficients ${B}_{n,\;m}$  for the nonlinear squeezed states $|\beta, \tilde{f}\rangle$ are
\begin{eqnarray}
\fl \quad \quad {B}_{n,\;m} = N^2_{\beta} \sum^{\infty}_{n, m = 0} \frac{{\beta^{*}}^m \beta^n}{2^{n+m} \;n!\;m!} \sqrt{\frac{(2n)!\;(2m)!}{(2n+2)!\;(2m+2)!\;(2n+3)!\;(2m+3)!}}.
\label{charc1}
\end{eqnarray}

To evaluate $C(\lambda, s)$, one can  derive the expression for $\langle 2m+3|D(\lambda)|2n+3\rangle$ as, \cite{spara, barn}  
\begin{eqnarray}
\langle 2m+3| D(\lambda) |2n+3 \rangle &=& e^{-\frac{|\lambda|^2}{2}}\sqrt{\frac{(2n)!}{(2m)!}} {\lambda^{*}}^{2n-2m}
L^{2n-2m}_{2m}(|\lambda|^2).
\label{s5}
\end{eqnarray}
where $ L^{2n-2m}_{2m}(|\lambda|^2)$ is an associated Laguerre polynomials \cite{book}. 

Using the expectation value (\ref{s5}) in (\ref{chars}) and then substituting the resultant expression in (\ref{spara}), we
arrive at
\begin{eqnarray}
F(z, s) = \frac{1}{\pi^2}\sum^{\infty}_{n,m=0} {B}_{n,\;m} \sqrt{\frac{(2 n)!}{(2 m)!}} \int \exp{\left[\frac{(s-1)}{2}|\lambda|^2 + \lambda^* z - \lambda z^*\right]} \nonumber \\
\qquad \qquad \quad \qquad \times {(\lambda^*)}^{2n-2m} L^{2n - 2m}_{2m}(|\lambda|^2)d^{2} \lambda.\;\;\;
\label{sspara2}
\end{eqnarray}

\begin{figure}[ht]
\centering
\includegraphics[width=0.6\linewidth]{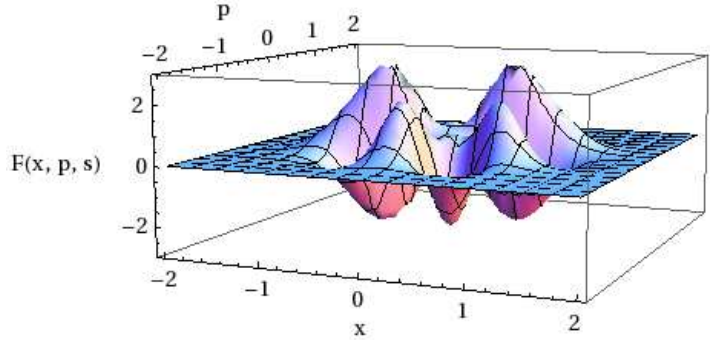}
\vspace{-0.1cm}
\caption{The plot of $s$-parameterized quasi probability distribution function corresponding to
 nonlinear squeezed states (\ref{lad26b}) for $s = 0.5$.} \label{ss_c1_wig}
\end{figure}

Evaluating the integral in (\ref{sspara2}), we find
\begin{eqnarray}
\fl \quad F(z, s) = \frac{2\exp{\left[\frac{2}{(s-1)}|z|^2\right]}}{\pi(1-s)}\sum^{\infty}_{n,m=0} {B}_{n,\;m} {(2 z^*)}^{2n-2m}\sqrt{\frac{(2n)!}{(2m)!}} \frac{(s+1)^{2 m}}{(s-1)^{2n}} \nonumber \\
\qquad \qquad \quad \qquad \times L^{2n-2m}_{2m}\left(\frac{4}{(1-s^2)} |z|^2 \right),
\label{sspara7}
\end{eqnarray}
where $ L^{2n-2m}_{2m}(|\lambda|^2)$ is an associated Laguerre polynomial.

We consider the value $s$ in-between $0 $ and $1$ and calculate a general quasi-probability distribution
function instead of investigating the special cases one by one, that is
(i) $s = 1$ (Glauber-Sudarshan $P$-function), (ii) $s = 0$ (Wigner function $W$) and
(iii) $s = -1$ (Husimi $Q$-function). Using (\ref{charc1}) in (\ref{sspara7}),
we numerically calculate $s$-parameterized quasi-probability distribution function, with $s = 0.5$
for the nonlinear squeezed states $|\beta, \tilde{f}\rangle$ (with $\beta = 2 + i2$)
and display the results in figure \ref{ss_c1_wig}  with
$z = x + i\; p$. The function $F(x, p, s)$ has negative values for the nonlinear squeezed states
$|\beta, \tilde{f} \rangle$. The results reveal the non-classical nature of the nonlinear squeezed
states.

\section{Conclusion}
In this paper, we have constructed nonlinear squeezed states for the generalized isotonic
oscillator by transforming the deformed ladder operators, which satisfy the deformed
oscillator algebra, suitably in such a way that they produce the Heisenberg algebra. We observed that
the transformation can be made in three different ways. While implementing this  we obtain
non-unitary squeezing operator in two cases and an unitary squeezing operator in the third case.
One of the two non-unitary squeezing operators produces the nonlinear squeezed states whereas
the other one fails to produce their dual pair. The unitary squeezing operator produces squeezed states only.
The non-classical nature of the nonlinear squeezed
states has been confirmed through the evaluation of photon number distribution, Mandel's parameter,
second-order correlation function and parameter $A_3$. Further, we have demonstrated that the nonlinear squeezed states
possess other non-classical properties as well, namely quadrature and amplitude-squared squeezing.
We have also analyzed the quadrature distribution  and $s$-parameterized quasi-probability function for the
nonlinear squeezed states which again confirmed the non-classical nature of these states.
The results summarized in this paper are all useful in the quantum entanglement perspective.

\end{document}